\documentclass[10pt]{article}
\usepackage{graphicx}
\begin{document}

\Large
\centerline{\bf A High Energy Electron and Photon Detector}
\centerline{\bf Simulation System}
\vskip 0.6cm
\centerline{\bf Srikanta Sinha}
\vskip 0.6cm

\normalsize
\centerline{3A, Sharda Royale Apt., 1 A Main, G M Palya, Bangalore-560 075,
INDIA}
\vskip 0.2cm
email: srikanta\_joy@yahoo.com
\vskip 0.4cm

\section{Abstract}
A detailed Monte-Carlo code has been developed from basic principles that 
simulates almost all of the basic photon and charged particle interactions.
The code is used to derive the response functions of a high energy photon  
detector to incident beams of photons of various energies.
The detector response matrice(DRM)s are calculated using this code. Deconvolution 
of an artificially generated spectrum is presented. 
\vskip 0.2cm
Keywords: Simulation, Detector response matrix, spectral deconvolution

\section{Introduction}
The objective of the present endeavour is to develop a simple and user-friendly 
detector simulation code.
An output pulse height spectrum is generated
by the present code that may be compared directly with a measured pulse
height spectrum which is the observable distribution. 

 The aim here is not to try to develop a simulation system as detailed and
sophisticated as say, the EGSnrc or GEANT4 software systems but one that has
all the basic physical processes incorporated within it and one which is, at the same
time, sufficiently simple so that it may be used with ease to compare results
of experiments where not very elaborate and complicated calibration systems
are available- hence the calibration data also has certain inherent uncertainties
and limitations due to effects (of say, surrounding materials etc.) that are neither
well understood nor properly taken care of. In other words the present simulation 
code will be very useful in applications where extremely accurate results
are not needed but a few percent accuracy will suffice.

    The approach taken here is to develop the entire detector simulation code 
starting from the very basic principles, generate artificial photon input
energy spectra, let these interact with the detector to give rise to output
pulse height spectra. Using the detector calibration data, the pulse height spectra
may be converted to equivalent output energy spectra.
Since the DRMs are calculated using the same program,
the output energy spectra can be deconvolved using these DRMs and the deconvolved
spectra can be compared with the original input photon spectra. This procedure 
establishes, to a large extent, the validity and accuracy of the entire
simulation code.

The simulation code is written in FORTRAN 77. The GNU compiler g77 is used to
compile the code. It runs under Redhat Linux 9.0 operating system. A typical run to simulate
a few thousand events for input photons say, 661 keV takes a few seconds
in a PENTIUM IV system. This gives the energy deposition spectrum. In order to
get the pulse height spectrum one has to use the photo-multiplier dynode multiplication
simulation procedure. This takes rather long time, typically more than two hours for
a few thosand events of 661 keV photons. The present code is nearly 3000 lines long (excluding the
data statements and the matrix inversion procedures). There are
plans to rewrite the entire code in FORTRAN 90/95.

\section{The Detector Geometry and Co-ordinate Systems}
In the present simulations the detector geometry assumed is a cylindrical one.
The center point of the detector is chosen to be the origin (O)
of the global coordinate (right-handed rectangular Cartesian) system.
The view axis of the detector is chosen to be the positive Z-axis of this co-ordinate 
system.
\begin{figure}[htp]
\includegraphics[height=8.5cm,width=14.5cm,angle=0]{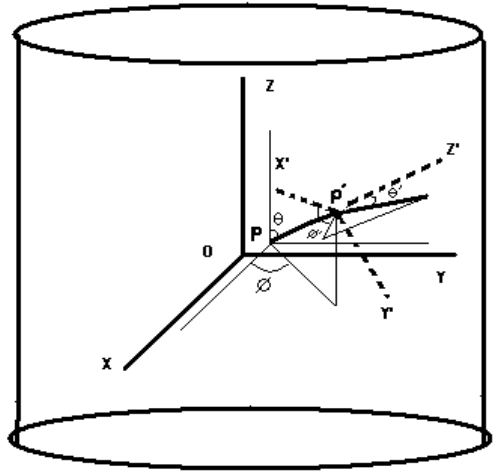}
\caption{The global (OXYZ) and local (P'X'Y'Z') co-ordinate systems in the cylindrical 
geometry.}
\end{figure}
\vskip 0.2cm

\subsection{Co-ordinate Transformations:}
Let a photon/particle start from a point P (x,y,z) and let its velocity (momentum)
vector has a polar angle of $\theta$ and azimuth of $\phi$ with respect to the
global co-ordinate system. If an interaction takes place at a geometrical 
distance $d_{int} (=d_{gm}/\rho)$ where $d_{gm}$ is the distance in $gm cm^{-2}$,
$\rho$ being the density of the medium, the co-ordinates of the interaction point
P'(x',y',z') with respect to the global frame may be written as (Fig.1)
\begin{equation}
x^{'}=x+d_{int} sin\theta cos\phi
\end{equation}
\begin{equation}
y^{'} = y + d_{int} sin\theta sin\phi
\end{equation}
and
\begin{equation}
z^{'} = z + d_{int} cos\theta
\end{equation}
 In the event of an interaction, let a product particle/photon be emitted at a
polar angle $\theta^{'}$ and azimuth $\phi^{'}$ where the $Z^{'}$-axis of the local co-ordinate
system is along the velocity (momentum) vector of the product particle/photon.
The values of $\theta^{'}$ and $\phi^{'}$ in the local co-ordinate system are transformed
back to the global system ($\Theta$, $\Phi$) as follows:
\begin{equation}
\left (\begin{array}{c}
 sin\Theta cos\Phi \\
 sin\Theta sin\Phi \\
 cos\Theta          \\
\end{array}
\right ) =
\left (\begin{array}{ccc}
 cos\theta cos\phi  & -sin\phi  & sin\theta cos\phi \\
 cos\theta sin\phi   & cos\phi   & sin\theta sin\phi \\
 -sin\theta          &   0       & cos\theta          \\
\end{array}
\right )
\left (\begin{array}{c}
 sin\theta^{'} cos\phi^{'}  \\
 sin\theta^{'} sin\phi^{'}  \\
 cos\theta{'}               \\
\end{array}
\right )
\end{equation}

 Procedures have also been developed for rectangular geometry. Trial runs
with these procedures gave encouraging results.
\section{Cross Sections}
  Cross sections for all the different types of interactions of photons and
charged particles are necessary to simulate these processes within the detector
material. Since the composition of the detector material is unknown a priori,
cross section tables for all elements should be available to the program. At
present the atomic cross section tables for only the following elements are available,
viz. Hydrogen, Carbon, Sodium, Argon, Iodine, Xenon and Cesium. Near the absorption 
edges the cross sections are calculated for many closely spaced energy values.
The XCOM program developed by Berger and Hubbell [1] is used for this purpose.
This program is used to calculate the photo-electric, coherent, incoherent 
and pair production cross-sections. 

    In the XCOM program, the coherent (Rayleigh) scattering cross sections are
calculated using a combination of the Thompson formula and relativistic Hartree-
Fock atomic form factors.

    In XCOM, the photo-electric cross sections are obtained following
the method of Scofield [2] upto 1.5 MeV of input photon energy. At higher energies
a semi-empirical formula [3] and the asymptotic high energy limit calculated by Pratt
[4] is used.

    In XCOM, the incoherent (Compton) scattering cross sections are obtained
using a combination of the Klein-Nishina formula and non-relativistic Hartree-
Fock incoherent scattering functions.

    For pair production, XCOM uses combinations of formulas from Bethe-Heitler
theory [5] alongwith other theoretical models to take into account screening, Coulomb
and radiative corrections. Please refer to ref.1 for the details of the XCOM
program.
Bremmstrahlung 
cross section is calculated separately.

   The total cross section for electron bremmstrahlung is taken as
\begin{equation}
\sigma_{rad} = 4\sigma_o [ln(183 Z^{-1/3}+1/18]
\end{equation}
 where $\sigma_o=\alpha Z^{2} r_e^{2}$, $\alpha$ is the fine structure constant, Z is
the atomic no. of the material. $r_e$ is the classical electron radius.
Photon and charged particle interaction cross section tables are prepared for a wide 
range of energies, viz. 1 keV to 1000 MeV.

Given the cross sections for the elements, the cross section for a composite
material (such as CsI) for any process may be calculated as
\begin{equation}
\sigma = \Sigma n_{i} \sigma_{i}
\end{equation}
where $n_{i}$ is the number of atoms per unit volume of the $i th$ element
constituting the material.
\subsection{The Radiation Length}
The radiation length $X_{0}$ is calculated as follows:
\begin{equation}
1/X_{0} = 4\alpha N/A Z(Z+1) r_{e}^{2} ln(183Z^{-1/3})
\end{equation}
Here $N$ is the Avogadro's number and $A$ is the mass number of the element.

   For a composite material the radiation length is given by the formula
\begin{equation}
1/X_{0} = p_{1}/X_{1} + p_{2}/X_{2}
\end{equation}
where $p_{1}$, $p_{2}$..., are the fractional weights of the various
components and $X_{1}$, $X_{2}$..., are the corresponding radiation lengths.
\section{Simulations of Different Processes}
\subsection{Photon Interactions}
   For photons four different types of interactions are considered. These are:
(i) Coherent scattering, (ii) Photo-electric absorption, (iii) Incoherent
scattering and (iv) Pair production. Atomic relaxations are considered in the
cases of Photo-electric absorption and Compton scattering.

 The geometrical distance of an interaction (d) from the point of incidence/previous 
interaction of the high energy photon is calculated from the equation
\begin{equation}
d = -\lambda\ln R
\end{equation}
where R is a uniform random number between 0 and 1.
All the uniform random numbers between 0 and 1 
are generated using the fortran RAND() function. $\lambda$ is the mean free path
(reciprocal of the total photon interaction cross section of the detection medium
at the given energy).
\begin{equation}
1/\lambda =\sigma_{coh.} +\sigma_{p.e.} +\sigma_{inc.} +\sigma_{pair}
\end{equation}
where $\sigma_{coh.}$, $\sigma_{p.e.}$, $\sigma_{inc.}$ and $\sigma_{pair}$ are
respectively the coherent, photo-electric, incoherent and pair production
cross sections. The value of the pair production cross section is, of course,
equal to zero below a photon energy of 1.022 MeV.
    The type ($K$) of the interaction (one of the four types mentioned above)
is determined using the relative (fractional) probability (cross section)
for the specific process using a Russian Roulette, i.e.
\begin{equation}
p_k =\sigma_k/\Sigma\sigma_k
\end{equation}
 $\sigma_k$ and $p_k$ being respectively the cross section and the relative probability 
for the process $K$.

   In the following we present the details of the algorithms used to simulate each of
the above mentioned interactions.
\subsection{Photo-electric Absorption:}
In the photo-electric interaction with the atom, the incoming photon
disappears and an electron is emitted whose energy equals the photon
energy minus the binding energy of the electron in the relevant atomic
shell.
\begin{equation}
 E = E_{p} - \phi
\end{equation}
where the symbols have their usual meanings. The atom then undergoes
relaxation through either fluorescent photon emission or non-radiative
transition (described later).
\subsection{Incoherent (Compton) Scattering:}
   To simulate the incoherent scattering process first the atomic species is
selected using a Russian Roulette (based on the relative incoherent scattering
cross sections of the different constituent atoms of the detector material).
Next, the relevant atomic shell is selected using the occupation numbers of
the different atomic shells, $Z_i/Z$.
   The scattering angle (polar) $\theta$ of the photon in the local co-ordinates
(the $Z^{'}$ axis of the local co-ordinate system points along the momentum vector
of the interacting photon)
is generated using the Klein-Nishina formula for the differential scattering cross
section
\begin{equation}
d\sigma/d\Omega = Z r_e^{2} (1/1+\alpha (1-x))^{2} (1+ x^{2} /2) (1+\alpha^{2} (1-x)^{2}/
(1+x^{2})[1+\alpha (1-x)])
\end{equation}
  Here $x=cos \theta$, $\theta$ being the polar angle of the scattered photon.
  $\alpha =1/137$ is the fine structure constant and $r_e$ is the classical electron radius.
  The von Neumann rejection technique is used for this purpose. Binding energy effects
are taken into account by considering only those energy transfers to the electron
that are larger than its binding energy. Doppler broadening effects are neglected.

     The energy of the scattered photon h$\nu^{'}$ is calculated
using the formula
\begin{equation}
h\nu^{'} = h\nu/[1+ (h\nu/m_o c^{2}) (1-cos\theta)]
\end{equation}
 $\nu$ being the frequency of the incident photon. $h$, $m_o$ and $c$ are respectively the
Planck's constant, the electron rest mass and the velocity of light.
The azimuth $\phi$ is uniformly generated between 0 and $2\pi$.
   
    The polar scattering angle $\theta_e$ (in the local system) of the recoil electron
is calculated as
\begin{equation}
\theta_e = tan^{-1} (1/(1+\alpha) tan (0.5\theta_p))
\end{equation}
where $\theta_p$ is the scattering angle of the photon and $\alpha = h\nu/m_o c^2$.
The azimuth $\phi^{'}$ of the recoil electron is equal to ($\pi-\phi_p$) while its
energy $E_e$ is given by
\begin{equation}
E_e = E  - E_p 
\end{equation}
\subsection{Coherent Scattering:}
The angular distribution $d\sigma_{coh}/d\Omega$ for coherent (Rayleigh) 
scattering is given approximately, by the distribution
function $d\sigma_{T}/d\Omega$ for classical Thomson scattering by an
electron,
\begin{equation}
d\sigma_{T}/d\Omega = r_{e}^{2}/2 (1+cos^{2}\theta)
\end{equation}
\subsection{Electron-Positron Pair-production}
      The screening parameter $\gamma$ is defined as
\begin{equation}
 \gamma = 100 (m_o c^{2} /E) [1/v(1-v)] Z^{-1/3}
\end{equation}
where
\begin{equation}
v = (E^{'}+m_o c^{2})/E
\end{equation}
      Here E is the energy of the photon that undergoes pair production into a
positron having total energy $E^{'}$ and an electron having total energy $E-E^{'}$.
$m_o$ is the electron rest mass, c the velocity of light and Z is the atomic number of the
target nucleus.
       The differential cross section for pair production $\Phi_{pair}(E,E^{'})$ depends on the value of the
screening parameter. The theoretical expression for $\Phi_{pair(E,E^{'})}$ is given by [6] as
\begin{equation}
\Phi_{pair}(E,E^{'}) dE^{'} = 4 \alpha N (Z^{2}/A) r_{e}^{2} (dE^{'}/E) G(E,v)
\end{equation}
       For different ranges of the value of the screening parameter, the following expressions
are used.
       No screening ($\gamma >> 1$):
\begin{equation}
G(E,v) = [v^{2}+(1-v)^{2}+(2/3)v(1-v)] [ln(2E/m_o c^{2}) v(1-v) -1/2]
\end{equation}
       Complete screening ($\gamma =0$):
\begin{equation}
G(E,V) = [v^{2}+(1-v)^{2}+(2/3) v(1-v)]ln (183Z^{-1/3})-(1/9) v(1-v)
\end{equation}
       Intermediate cases ($0<\gamma< 2$):
\begin{equation}
G(E,v) = [v^{2}+(1-v)^{2}] [f_{1}(\gamma)/4-(1/3) lnZ]
\end{equation}
       ($2<\gamma< 15$):
\begin{equation}
G(E,v) = [v^{2}+(1-v)^{2}+(2/3) v(1-v)] [ln(2E/m_o c^{2}) v(1-v)-1/2-c(\gamma)]
\end{equation}
       The functions $f_{1}(\gamma)$, $f_{2}(\gamma)$, and $c(\gamma)$ are the 
same that enter in the expressions for the radiation probabilities mentioned
in subsection 7.1.
      For very small energy, k very near to $2m_o c^{2}$, Hough's approximate formula
\begin{equation}
\Phi(E_{+})dE_{+} =8/3\Phi_{0} (k-2m_o c^{2}/k)^{3} z dE_{+}
\end{equation}
is used, where
\begin{equation}
z=2\sqrt x(1-x)
\end{equation}
and x is given by
\begin{equation}
x=E_{+}-m_o c^{2}/k-2m_o c^{2}
\end{equation}
  The angle of the positron (electron) with respect to the momentum vector of the high
energy photon (Z' axis of the local system) is given by
\begin{equation}
\theta = m_o c^{2}/E_\pm
\end{equation}
\subsection{Muon Pair Production:}
The formulae given in the previous subsection may be used to simulate
$\mu^{+}-\mu^{-}$ pairs although the threshold for this process is
equal to $206$ times the threshold for the corresponding process
for $e^{+}-e^{-}$ pair, i.e. 210.532 MeV.
\section{Atomic Relaxations:}
 In the case of photo-electric absorption and also in the case of incoherent
scattering, a vacancy is created in the electronic shell with which the photon 
interacts. This vacancy is filled by electrons from higher energy shells and the
excitation energy is released in the form of either (i) one or more fluorescent
photons, (ii) an Auger electron, or (iii) a Coster-Kronig electron. The relative
probabilities for these three types of relaxation processes are different for
different electronic shells. In the present work the Coster-Kronig process is
neglected.
\subsection{Fluorescence}
The energy of the fluoresnce photon is equal to the difference in the binding
energies of the initial and final states of the atom.
The fluorescent X-ray photon is emitted isotropically. Its polar angle ($\theta$)
and azimuth ($\phi$) in the local system are calculated as follows:
\begin{equation}
\theta=cos^{-1}(2R_1-1)
\end{equation}
and
\begin{equation}
\phi=2\pi R_2
\end{equation}
$R_1$ and $R_2$ being two different random numbers that are distributed uniformly
between 0 and 1. The values of $\theta$ and $\phi$ are transformed back to the 
global co-ordinate system using eqns.(1) and (2).

      The fluorescent photon may, either (i) escape the detection medium (the
location of its interaction being outside the boundary of the detector and in that 
case its energy is treated as lost resulting in an energy deposit that is equal to 
the input photon energy minus the fluorescent photon energy. This kind of events
give rise to the characteristic X-ray escape peak), or, (ii) it may
interact photo-electrically with a higher shell in which case either its total 
energy or a part of its energy (in the case of escape of a higher shell
fluorescent photon) is deposited.
\subsection{Auger Electron Emission}
The Auger electron emission (non-radiative electron emission) is considered
only for the K and L shells. The energy(ies) of the Auger electron(s) is completely
absorbed within the detection medium and contribute to the energy deposited.
\section{Charged Particle Interactions:}
   For electrons, bremmstrahlung and for positrons, bremmstrahlung and pair
annihilation processes are considered. For electrons and positrons multiple
Coulomb scattering is considered in an approximate manner.
\subsection{Bremmstrahlung}
      The Bremmstrahlung process is essentially similar to the pair production
phenomenon.

  The differential cross section (probability) for the radiation process,
$\Phi_{rad}(E,E^{'})$ depends on the value of the screening parameter which
is defined as
\begin{equation}
\gamma = 100 (m_o c^{2}/U) [v/(1-v)] Z^{-1/3}
\end{equation}
  Here,
\begin{equation}
U = E + m_o c^{2}
\end{equation}
and
\begin{equation}
v = E^{'}/U
\end{equation}
      The expression for $\Phi_{rad}(E,E^{'})$ is given by ref.6 as
\begin{equation}
\Phi_{rad}(E,E^{'}) dE^{'} = 4\alpha (N/A) Z^{2} r_{e}^{2} (dE^{'}/E^{'}) F(U,v)
\end{equation}

      For different ranges of the value of $\gamma$, the differential cross section 
is given by the following formulas:
      No screening ($\gamma >> 1$ ):
\begin{equation}
F(U,v) = [1+(1-v)^{2}-(2/3)(1-v)] [ln(2U/m_o c^{2}) (1-v)/v-1/2]
\end{equation}
     Complete screening ($\gamma =0$ ):
\begin{equation}
F(U,v) = [1+(1-v)^{2}-(2/3) (1-v)] ln(183Z^{-1/3}) +1/9 (1-v)
\end{equation}
     Intermediate cases, ($0<\gamma< 2$ ):
\begin{equation}
F(U,v) = [1+(1-v)^{2}] [f_{1}(\gamma)/4-(1/3) lnZ] -(2/3) (1-v) [f_{2}(\gamma)/4-(1/3) lnZ]
\end{equation}
    ($2<\gamma<15$ ):
\begin{equation}
F(U,v) = [1+(1-v)^{2}-(2/3) (1-v)] [ln(2U/m_o c^{2}) (1-v)/v-1/2-c(\gamma)^{1/2}]
\end{equation}

  The average angle of the emitted photon with respect to the momentum vector of the high
energy electron ($Z^{'}$ axis of the local system) is given by
\begin{equation}
\theta_{0} = m_o c^{2}/E_{0}
\end{equation}
\subsection{$e^{+} e^{-}$ Annihilation}
   Only electron-positron pair annihilation at rest is presently considered.
The energy of each annihilation photon is equal to $0.511$ MeV.
The direction of the second photon is taken to be exactly opposite to the
direction of the first photon which is sampled from a distribution that is
isotropically distributed.
\subsection{Knock-On ($\delta$ ray) production}
High energy electrons often produce electrons of energies that are comparable to the
energies of the primary electrons. This process is simulated using the formula [7]
\begin{equation}
P(E)dE = W dE/E^{2}
\end{equation}
where $P(E) dE$ is the probability of an electron receiving the energy $E$.
$W$ is a constant that depends on $Z$, $A$; the atomic number and the mass number
of the medium in addition to the velocity $\beta$ of the primary electron. 
\subsection{Multiple Coulomb Scattering}
      The well-known Moliere theory of Coulomb scattering is used in order to take into
account the scattering of light charged particles (electrons and positrons) in the Coulomb fields
of both atomic nuclei and atomic electrons.
     
      For this the energy domain is divided into two intervals. The first interval
includes charged particles whose energies lie between 1 keV and 1 MeV. The second interval 
includes the energy region between 1 MeV and 1 GeV. These two energy intervals are
again subdivided each into 30 logarithmic energy bins (following the treatment used by
Vatcha [8]). For charged particles having energies above 1 GeV, only a gaussian having
a width equal to $(E_{s}/E\beta)\sqrt(t)$,
(where $E$ is the energy of the particle, $\beta=v/c$, v being its velocity and
$t=x/X_{0}$ is the thickness of the slab of material in terms of its radiation
length) is considered. The constant $E_{s}= 21$ MeV. 

      Bethe [9] has expressed Moliere's theory
of multiple Coulomb scattering in a form which is easy to simulate. Using the first two 
terms of his equation () together with a correction for solid angle the frequency
function for scattered angle is
\begin{equation}
f(\Theta) d\Theta = (sin\Theta/\Theta)^{1/2} [f^{(0)}(\phi) + 1/B f^{(1)}(\phi)] \phi d\phi
\end{equation}
 Here $\phi$ is the reduced angle of $\Theta$ which is defined as
\begin{equation}
\phi = \Theta/\chi_{c} B^{1/2}
\end{equation}
where
\begin{equation}
\chi_{c}^{2} = 0.157 Z(Z+1) t/A (pv)^{2}
\end{equation}
where $\it pv$ is in MeV, $t$ is in grams per square centimeter, and $A$ is the atomic
weight of the material.
    The parameter B is defined by the transcendental equation
\begin{equation}
B - ln B = B^{'}
\end{equation}
where
\begin{equation}
% B^{'} = ln (\chi_{c}^{2}/1.167 \chi_{a}^{2})
B^{'} = 2 ln (\chi_{c}/\chi_{a}) +1 -2C
\end{equation}
where $C=0.577...$ is Euler's constant and $\chi_{a}$ is an angle which
depends on the screening of the atomic field. It is defined as
\begin{equation}
% \chi_{a}^{2}=6.8X10^{-5} Z^{2/3} [1.13+3.76 (\alpha Z)^{2}] 1/\beta^{2}E^{2}
\chi_{a}^{2} = (\lambda/5.561a_{0}Z^{-1/3})^{2} (1.13+3.76\alpha^{2})
\end{equation}
with
\begin{equation}
\alpha = 2\pi zZ^{2}/hv
\end{equation}
The functions $f^{(0)}(\phi)$ and $f^{(1)}(\phi)$ are defined as
\begin{equation}
f^{(0)}(\phi) = 2 e^{-\phi^{2}}
\end{equation}
and
\begin{equation}
f^{(1)}(\phi) =f^{(0)}(\phi) (\phi^{2}-1) (Ei(\phi^{2})-ln(\phi^{2})-2 (1-f^{(0)}(\phi))
\end{equation}
where $Ei(x)$ is defined as follows:
\begin{equation}
Ei(x) = -\int_{-x}^{\inf} e^{-t}/t dt = \int_{-\inf}^{x} e^{t}/t dt
\end{equation}
For a given electron energy a slab thickness (proportional to the energy
of the electron) of the material is chosen. The percentage probability of
scattering is selected using a uniform random number between 0 and 1.
The angle of scattering is selected from the table using reverse interpolation [10].
\subsection{Ionisation Loss and Particle Ranges}
For the ionisation loss the expression given by Bethe and Bloch in ref.7 is used.
\begin{equation}
dE/dx = (1/\rho) dE/dX
\end{equation}
Here $x=X\rho$ is in $gms cm^{-2}$ while $X$ is in cms.
\begin{equation}
dE/dX = -K (Z/A) (\rho/\beta^{2}) [ln 2m_o c^{2}\beta^{2}E_{M}/I^{2}(1-\beta^{2})-2\beta^{2}]
\end{equation}
where
\begin{equation}
K = 2\pi N z^{2}e^{4}/m_o c^{2}
\end{equation}
Here $N$ is the Avogadro number, $m_o$ and $e$ are mass of the electron and its charge,
$Z$, $A$ and $\rho$ are the atomic number and the mass number and the density of the
medium, respectively, and $I$ is its effective ionization potential; $z$ is the charge
and $\beta$ the velocity (in units of $c$, the speed of light) of the incident particle.
 
 $E_{M}$ represents the maximum possible energy transfer in an interaction and is
given by
\begin{equation}
E_{M} = 2m_o c^{2}\beta^{2}/(1-\beta^{2})
\end{equation}
For energies of electrons ($\delta$ rays) an approximate formula for the practical
range, in $g cm^{-2}$ is used (ref.7)
\begin{equation}
R_{p} = 0.71 E^{1.72}
\end{equation}
where the energy of the electron ($E$) is expressed in MeV.
\subsection{Emission of Scintillation Photons}
The number of scintillation photons (integrated over the spectral band) produced in the 
scintillator is obtained simply by dividing the energy deposied by a constant that 
gives the average energy reqired to produce a scintillation photon. The value of this
constant is 0.02564 eV for CsI [11]. Since the number of scintillation photons is 
always an integer, 0.5 is added to the calculated value and the result is truncated.
\subsection{Emission of Cerenkov Photons}
  In case the charged particles produced in any of the interactions happen to
traverse regions that consist of transparent media (such as air, water, plastics etc.)
and the velocity of the particles exceed the velocity of light in those
media, Cerenkov photons are emitted.

    The angle of emission $\theta_{C}$ of the Cerenkov photons is given by the equation
\begin{equation}
cos \theta_{C} = 1/n\beta
\end{equation}
where $\beta$ is equal to $v/c$, $v$ being the velocity of the particle, while
$n$ represents the refractive index of the local medium.
     The number of Cerenkov photons $dN$ emitted in the track length element $dl$ in the
wavelength interval $\Delta\lambda$ is given by the equation
\begin{equation}
dN = 2\pi\alpha dl (1-1/n^{2}\beta^{2}) \Delta\lambda
\end{equation}
The spectral distribution of Cerenkov photons is sampled using the probability
distribution
\begin{equation}
P(\lambda)d\lambda = (1.0/\lambda^{2})d\lambda
\end{equation}
\subsection{Charge Multiplication at the Photomultiplier:}
 The scintillation photons incident on the photo-cathode ($N_i$) of the photo
multiplier (PMT) are converted into a photo-current (number of photo-electrons,$N_e$)
\begin{equation}
N_e=<\eta>N_i
\end{equation}
where $<\eta>$, the average spectral efficiency of the photo-cathode is 
taken to be equal to $~0.2$. 
$<\eta>$ depends on the spectral response of the particular photocathode type. 
Strictly speaking,
\begin{equation}
N_e=\int N_i(\lambda)\eta(\lambda)/\int \eta(\lambda)
\end{equation} 
where $N_i$($\lambda$) is the number of scintillation photons having wavelength 
$\lambda$ and $\eta$($\lambda$) is the corresponding spectral efficiency.

     The photo-electrons are multiplied by the successive dynodes (8 to 10 in number)
through secondary emission process.
     
    The energy resolution of the detector is determined largely by the statistical
fluctuations in the minimum number of particles (electrons) in the dynode chain, i.e.
the number of photo-electrons. The charge multiplication at the dynodes is simulated
in the following manner.

    Using the knowledge of the total PMT high voltage and its distribution between 
different dynodes (usually all the dynode voltages are equal except the first one),
from the PMT data book the average secondary electron multiplication factors are
obtained. Since the electron multiplication is a Poisson process, in each succesive
stage this multiplication is simulated using a previously calculated table of
Poisson probabilities for obtaining an integral number of secondary electrons
for a given average value.
\subsection{Folding the Radial Response in the case of scintillation detectors}
It has been observed that in scintillation detectors, the photomultiplier tube
light collection efficiency reduces to $85\%$ near the edge of the crystal for a
given energy deposition, as compared to the same energy deposited at the center
of the crystal. This radial radial response function (a quadratic fit to light
collection vs. radial distance of the point of energy deposition) is used to
obtain the amount of light incident at the photo-cathode of the PMT.
\subsection{Charge Integration and Pulse Height Spectrum}
The charge output at the anode of the PMT is integrated into a charge sensitive
pre-amplifier (CSPA) having a charge gain equal to $0.025X10^{12}$ Volts/Coulomb,
thus producing a voltage pulse as output. This is amplified using a linear
amplifier of gain equal to 5.
\subsection{Calibrating the Detector}
 The detector is calibrated as follows:
 The response functions (the pulse height spectra) are simulated for two monoenergetic
incident photons (say, 100 keV and 600 keV) using a large (say, 10 000) number of
incident photons. The positions of the photo-peaks (full energy peaks) are determined
accurately using very fine binnings.
   A straight line fit is obtained with the pulse height as a function of the input
photon energy
\begin{equation}
 h = m E + b
\end{equation}
h being the pulse height in volts and E being the photon energy in keV.
The calculated values of m and b are used to convert output pulse heights into
corresponding energy values to form the output energy spectra.
\section{Generation of Artificial Photon Input Spectrum and Spectral Deconvolution:}
   In order to check the correctness of the calculated DRMs it is necessary to
generate artificial photon energy spectra, let them interact with the detector 
and obtain output pulse height spectra. These pulse height spectra should then 
be deconvolved back into photon energy spectra using the calculated DRMs. These
deconvolved photon spectra should be compared with the original photon spectra
using say, $\chi^{2}$ tests. This procedure would then establish the correctness 
of the entire detector simulation code.
\subsubsection{Power-Law Photon Energy Spectrum}
  A power-law (differential) photon energy spectrum having the form $AE^{-\gamma}$
where A is the normalisation constant, E the energy of the photon and $\gamma$
is the spectral (differential) index, is generated using the inverse transform 
method as follows:
\begin{equation}
E=(KR+k_1)^{1/(1-\gamma)}
\end{equation}
 where $K=E_2^{(1-\gamma)}-k_1$ and $k_1=E_1^{(1-\gamma)}$. $E_1$ and $E_2$ are
respectively the minimum and maximum limits of the spectral energy band. $R$ is
a random number uniformly distrubuted between 0 and 1.

An input spectrum of photons (power law) is simulated. The corresponding output
pulse height spectrum is converted to an equivalent energy spectrum (using the
artificial calibration data). Let $\bf P$ be the vector that represents the
output pulse height spectrum (rebinned) and $\bf R$ the detector response
matrix. If the true input photon energy spectrum is denoted by the vector $\bf S$, then
\begin{equation}
\bf
 P_i = \Sigma R_{ij} S_{j}
\end{equation}
where $\bf P_i$ is the detected counts in the $i$th energy channel, $\bf S_j$ the number of
input photons in the $j$th energy channel and $\bf R_{ij}$ is the $ij$th element of the
DRM.
   The true incident photon spectrum $\bf S$ is obtained by the matrix inversion procedure.
\begin{equation}
\bf
 S = R^{-1} P
\end{equation}
where $\bf R^{-1}$ is the inverse of the detector response matrix (historically,
the first such detector response matrix was calculated by J. H. Hubbell [12]).  
\subsection{Simulation of Gamma Ray Burst (GRB) Spectrum}
The energy spectrum of Gamma Ray Burst sources are best described by the Band [13],[14]
spectrum. A typical Band spectrum is given as
\begin{equation}
N(E) = A (E/100)^{\alpha} exp(-E/E_{0})
\end{equation}
for
\begin{equation}
E <= (\alpha-\beta)E_{0}
\end{equation}
and
\begin{equation}
N(E) = A (\alpha-\beta) (E_{0}/100)^{\alpha-\beta} (E/100)^{\beta}
\end{equation}
for
%\begin{equation}
\begin{displaymath}E > (\alpha-\beta)E_{0}\end{displaymath}
%\end{equation}
\section{Gamma Ray Applications}
  In the case of gamma ray studies, the detector (e.g. a CsI scintillator) is usually shielded
using a plastic scitillator that is used as a veto counter to remove the charged
particle background. Usually there is a thin window (say, of Alumnium) in front of the 
scintillator. In addition a metal casing is used to cover the detector.
Gamma rays interact in all of these materials and produce either coherently
scattered gamma ray photons having energies equal to that of the primary, or,
they may produce incoherently scattered photons of lower energies or they may also 
produce annihilation gamma rays. These secondary photons have finite chances of
entering the actual detection volume (in this case the CsI crystal). Sometimes they
may be scattered from another material and enter the scitillator though the probabilities
of such tertiary and higher order scatterings are small.

     These effects are taken care of in the following manner. Nested volumes of materials
are considered where the actual detection volume (in this case CsI) is designated
as the volume of zeroth order. The next set of volumes (composition will in general 
be very different) that enclose the zeroth volume is called the volumes of first
order and so on. When a primary gamma ray is incident on any of the volumes, its
history is traced until either it misses or hits the zeroth volume. If it misses
the zeroth volume it is rejected while in the other case it is considered as a valid
incident photon of degraded energy. Development of this procedure is yet to be
completed.
\section{Results}
In the following the results of simulations for a Cesium Iodide scintillation detector
(3 inch diameter and 0.5 inch thickness) are presented.
\subsection{Detector Efficiency}
   The efficiency of a Cesium Iodide scintillation detector has been estimated from
the simulations. These estimated values of the detector efficiency are plotted in
Fig.2.  
\begin{figure}[htp]
\includegraphics[height=5cm,width=6.5cm,angle=0]{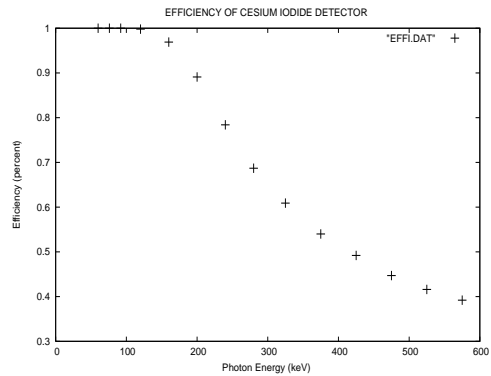}
\caption{The efficiency of the CsI detector (3 inch diameter and 0.5 inch thickness)
 is plotted as a function of energy.}
\end{figure}
\vskip 0.5cm
\subsection{Energy Resolution of the Detector}
  From the simulations the energy resolutions of a Cesium Iodide scintillation detector 
has been derived at various energy values. The energy resolution is obtained by dividing the
full width at half maximum (FWHM) of the photo-peak (full energy peak) by the energy of the
photo peak.
In Fig.3 the estimated energy resolutions are plotted against the energy of the incident
high energy photon.
\begin{figure}[htp]
\includegraphics[height=5cm,width=6.5cm,angle=0]{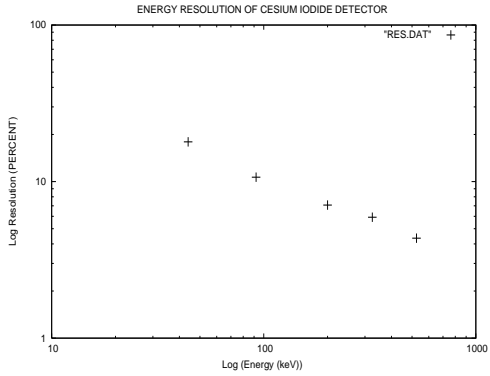}
\caption{The energy resolution of the CsI detector is plotted as a function of
energy. The $E^{-1/2}$ dependence of the energy resolution is clearly evident
in the figure.} 
\end{figure}
\vskip 0.2cm
\subsection{Sample Response Functions}
 In Fig.4 a sample response function is shown. The energy of the incident photons are
each equal to $70$ keV.
In Fig.4 the peak near 20 keV is due to the escape of the characteristic X-ray fluorescence
photons. For Cesium and Iodine, the energies of these photons are respectively 35.98 and 
33.17 keV. 
\begin{figure}[htp]
\includegraphics[height=5cm,width=6.5cm,angle=0]{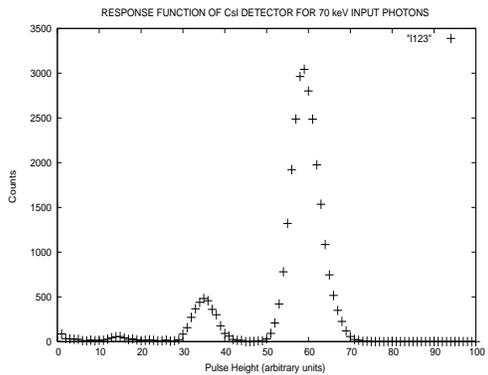}
\caption{Simulated response function of the CsI scintillator to an input beam of
$70$ keV photons. The K X-ray escape peak is clearly seen.}
\end{figure}
In Fig.5 the response function of the same detector for 525 keV input photons are
shown.
\vskip 0.2cm
\begin{figure}[htp]
\includegraphics[height=5cm,width=6.5cm,angle=0]{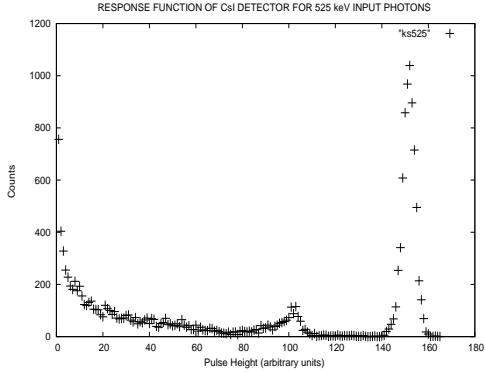}
\caption{Simulated response function of the CsI scintillator to an input beam of
$525$ keV photons. The Compton edge and the Compton continuum are clearly visible.}
\end{figure}
\vskip 0.2cm
\subsection{Response Matrix and Spectral Deconvolution}
The response functions of the CsI detector are calculated for the following
16 different energy values, viz. 28, 44, 60, 76, 92, 120, 160, 200, 240,
280, 325, 375, 425, 475, 525 and 575 keVs. These response functions are
binned into the following energy intervals, viz. 12-30, 30-49, 49-68,
68-87, 87-106, 106-152, 152-199, 199-246, 246-293, 293-340, 340-398,
398-457, 457-516, 516-574 and 574-633 keV respectively. Thus, the response
matrix (DRM) of the CsI detector is obtained.

 A power-law (differential index equal to -1.38) photon spectrum is
generated using the algorithm described in section $7.0.1$. A total of 9 936
photons having their energies between 20 keV and 600 keV are generated.
In addition, 70 photons having energy equal to 511 keV each are generated.
These photons are made incident on the CsI scintillator at an angle of
1 degree relative to the view axis. These photons interact within the
scintillation crystal and give the output spectrum depicted in fig.6.
This output spectrum is obtained by binning the output energy spectrum
using the energy intervals described in the previous paragraph.
\begin{figure}[htp]
\includegraphics[height=5cm,width=6.5cm,angle=0]{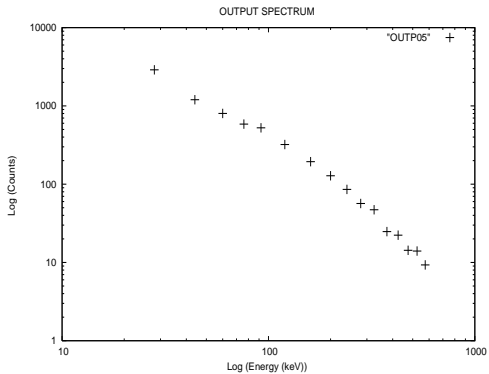}
\caption{Simulated output spectrum from the detector for an input power
law photon spectrum (differential index equal to $-1.38$. At higher energies
the counts decrease due to the gradually decreasing efficiency of photon
detection.} 
\end{figure}
\vskip 0.2cm
The inverse of the DRM is calculated using well established procedures [15].
The result of deconvolution of the output energy spectrum of fig.6 with
this inverse of the DRM is shown in the following figure.
\begin{figure}[htp]
\includegraphics[height=5cm,width=6.5cm,angle=0]{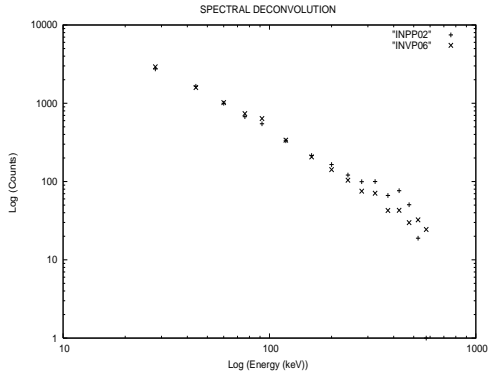}
\caption{The deconvolved energy spectrum (crosses) is compared with the
input photon spectrum (plus signs). The two spectra agree quite well at low and medium
energies but seem not to match very well at the high energy end.}
\end{figure}
\vskip 0.2cm
\section{Results of Simulations of Other Interaction Processes}
Pair production at high energies is simulated using the formulae
given in section 4.5. The results are shown in the figure below.
In Fig.8 the fractional energy ($v$) of the positron is plotted
along the abscissa and the quantity $E\phi_{pair}(E,E^{'})$ is
plotted along the ordinate. Here $\phi_{pair}(E,E^{'}) = X_{0}\Phi_{pair}(E,E^{'})$
is the differential pair production probability per radiation length.
$\Phi_{pair}(E,E^{'})$ is the differential pair production probability of
production of a positron having energy $E^{'}$ by an incident photon
having energy $E$ in $1 gm cm^{-2}$ thickness of the material.
\begin{figure}[htp]
\includegraphics[height=5cm,width=6.5cm,angle=0]{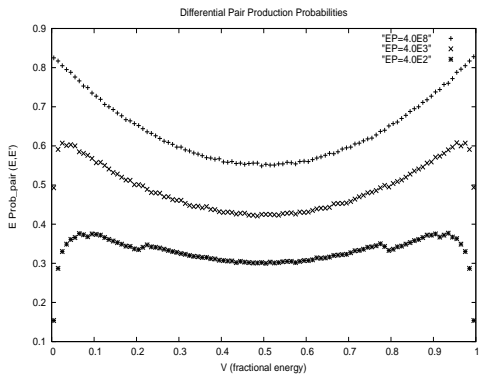}
\caption{Simulated differential pair production probabilities at high energies.
The values (inset) give the energies of the input photons in terms of the
electron mass.}
\end{figure}
\vskip 0.2cm
The pair production at low energies are simulated using the Hough's 
formulae given in section 4.5. The results of one such simulation is shown
in the following figure.
\begin{figure}[htp]
\includegraphics[height=5cm,width=6.5cm,angle=0]{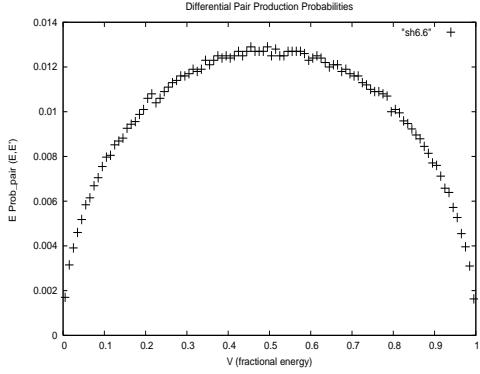}
\caption{Simulated differential pair production probabilities at low energies.
The value (inset) give the energy of the input photons in terms of the
electron rest mass.}
\end{figure}
\vskip 0.2cm
The electron/positron bremmstrahlung processes at high energies are
simulated using the formulae given in section 6.1. The results of these
simulations are given in fig.10. Here again the fractional energy ($v$)
of the emitted photon is plotted along the abscissa and the quantity
$E^{'}\phi_{rad}(E,E')$, where $\phi_{rad}(E,E^{'}) = X_{0}\Phi_{rad}(E,E^{'})$, 
is the differential radiation probility per radiation length.
$\Phi_{pair}(E,E^{'})$ is the differential radiation probability of
production of a photon having energy $E^{'}$ by an incident electron 
having energy $E$ in $1 gm cm^{-2}$ thickness of the material.
\begin{figure}[htp]
\includegraphics[height=5cm,width=6.5cm,angle=0]{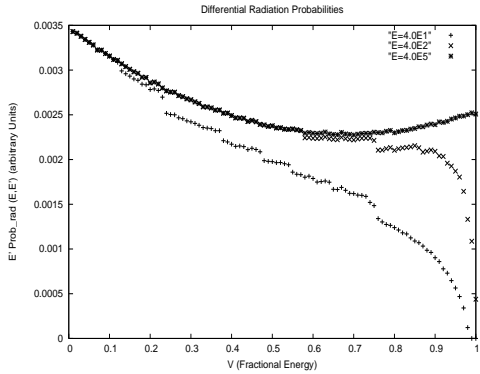}
\caption{Simulated differential radiation probabilities at high energies.
The values (inset) give the energies of the input electrons in terms of
the electron rest mass.}
\end{figure}
\vskip 0.2cm
Knock-on (delta ray) productions by a 10 MeV electron are simulated using the
procedure described in section 6.3. The results are shown in fig.11.
\begin{figure}[htp]
\includegraphics[height=5cm,width=6.5cm,angle=0]{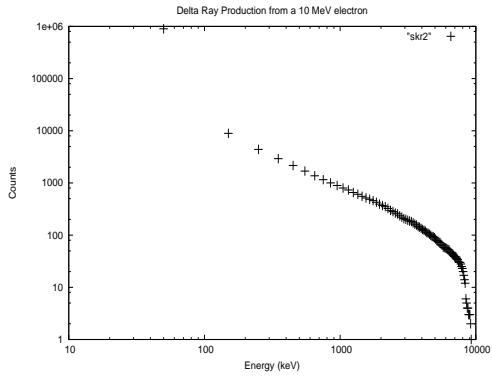}
\caption{Simulated probability distribution for the production of delta rays
(knock-on electrons) produced by an incident electron of energy $10$ MeV.} 
\end{figure}
\vskip 0.2cm
Cerenkov photon emission by high energy electrons/positrons are 
simulated using the formulae given in section.6.7. The simulated
wavelength spectrum of Cerenkov photons in the wavelength band of $300 nm$
to $650 nm$ is shown in the following figure.
\begin{figure}[htp]
\includegraphics[height=5cm,width=6.5cm,angle=0]{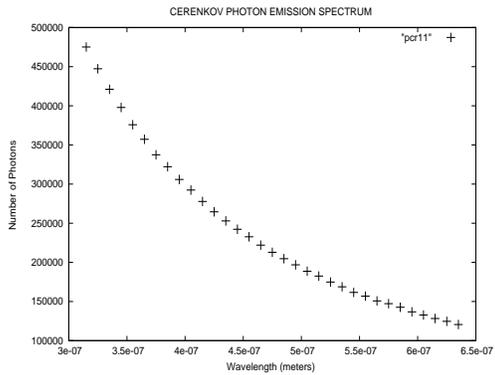}
\caption{Simulated wavelength spectrum of cerenkov photons emitted by an
electron/positron within the wavelength range of $300 nm$ to $650 nm$.}
\end{figure}
\vskip 0.2cm
\section{Simulations of Photon Energy Spectra}
Different sources emit different types of high energy photon
spectra. It is necessary, therefore, to be able to simulate
different types of photon energy spectra. In the following
two different types of spectra, viz. (a) a power-law
energy spectrum (large number of photons- the total being equal
to 9 million) and (b) a smoothly-broken power-law (Band Spectrum)
spectrum are shown.
\begin{figure}[htp]
\includegraphics[height=5cm,width=6.5cm,angle=0]{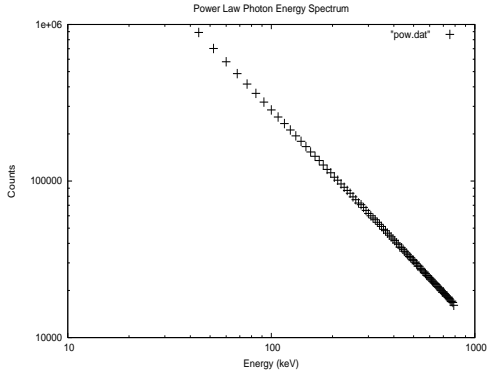}
\caption{Simulated power-law energy spectrum of photons having a
differential energy index of $-1.38$.}
\end{figure}
\vskip 0.2cm
\begin{figure}[htp]
\includegraphics[height=5cm,width=6.5cm,angle=0]{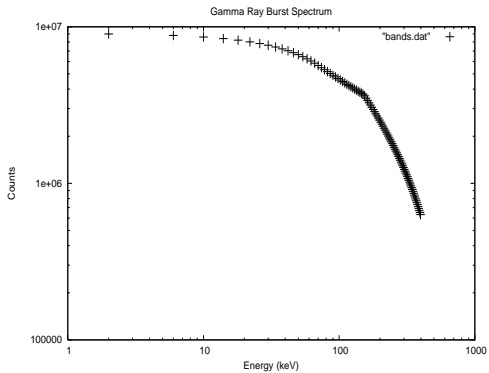}
\caption{Simulated energy spectrum (Band spectrum) of a hypothetical
gamma ray burst. The parameter values are: (i) $A=1.0E4$, (ii)
$\alpha= -1.0$, (iii) $\beta =-2.0$ and (iv) $E_{0} =150.0 keV$.}
\end{figure}
\vskip 0.2cm
\section{Discussion}
 It is possible to incorporate any detector geometry (say, rectangular geometry)
in the present simulation
code and calculate the response of that detector. Also, 
the response functions of other types 
of detectors, for example, gas (Argon or Xenon) filled 
multi-wire proportional chambers (MWPCs) such as those
used in X-ray astronomy experiments may be obtained.

In the present simulations the differential cross sections used for
the coherent scattering and the incoherent scattering processes
are respectively the classical Thomson cross section and the
Klein-Nishina formula. Strictly speaking, in the case of coherent scattering
the square of the atomic form factor $F(q,Z)$ should be multiplied with
the Thomson cross section while in the case of incoherent scattering the
incoherent scattering function $S(x,Z)$ should be multiplied with the
Klein-Nishina differential cross section. These changes will be 
introduced in a later version of the code.

All the different modes
of atomic relaxations are yet to be incorporated in the present code. 
Pair annihilation in flight and triplet production
are also to be introduced.

There are some small irregularities and discontinuities 
in the plots of the differential probabilities
for pair production and bremmstrahlung (fig.5 and fig.7). Presumably these have
resulted due to the fact that somewhat coarse interpolations of the
three screening functions ($f_{1}(\gamma)$, $f_{2}(\gamma)$, and
$c(\gamma)$) have been used in the present simulations.

There is minor disagreement of the deconvolved spectrum with the input
photon spectrum (Fig.5) at the high energy end. This is due to the fact 
that the number of events in this energy range are small and the response
functions at the higher energies are poorly determined (due to the
insuffiency of the number of events generated to calculate the 
response functions).
\section{Conclusions}
A high energy photon detector simulation code developed from the first
principles has been described. The results obtained using this code
are quite encouraging. This code gives an output
pulse height spectrum, instead of an energy loss spectrum.
This simulation code is being developed
further to incorporate all the possible interaction types of photons
and particles. It should be possible to extend this code to simulate
extensive air showers (EAS) in the atmosphere produced by very high
energy (VHE) and ultra high energy (UHE) gamma rays and electrons
and also to study the cerenkov radiation produced in these cascades.
\section{Acknowledgements}
 I would like to express my deep sense of gratitude to all my revered teachers.

This research article is dedicated to the memory of late Prof. John H. Hubbell.
I am greatly indebted to him and his collaborators at NIST for
kindly supplying me the XCOM program. Prof. Hubbell had also given me a lot of
relevant literature on photon cross sections as well as other material that
were very helpful to my understanding. It is my bad luck that I could not
finish this work before he passed away. 

I thank Prof. Kevin Hurley and Prof. Fabio Sauli for their kind
suggestions to improve the manuscript.
Thanks are also due to Dr. Varsha Chitnis for her kind help.
I thank my family for their patience and support.

\end{document}